\documentclass[twocolumn,showpacs,preprintnumbers,amsmath,amssymb,A4paper]{revtex4}

\usepackage[dvips]{graphicx}
\usepackage{dcolumn}
\usepackage{bm}

\begin{document}
\newcommand{\kp}{{\bf k$\cdot$p}\ }
\newcommand{\Pp}{{\bf P$\cdot$p}\ }

\preprint{APS/123-QED}
\title{Temperature dependence of the electron spin $\textbf{g}$ factor in CdTe and InP}

\author{Pawel Pfeffer$^*$ and Wlodek Zawadzki}
 \affiliation{Institute of Physics, Polish Academy of Sciences\\
 Al.Lotnikow 32/46, 02--668 Warsaw, Poland\footnotetext{$^*$ e-mail address: pfeff@ifpan.edu.pl}\\}
\date{\today}

\begin{abstract}
Temperature dependence of the electron spin $g$ factors in bulk CdTe and InP
are calculated and compared with experiment. It is assumed that the only modification of the band structure
related to temperature is a dilatation change in the fundamental energy
gap. The dilatation changes of fundamental gaps are calculated for both materials
using available experimental data. Computations of the band structures in
the presence of a magnetic field are carried out employing five-level {\Pp}
model appropriate for medium-gap semiconductors. In particular, the model
takes into account spin splitting due to bulk inversion asymmetry
(BIA) of the materials. The resulting theoretical effective masses and $g$
factors increase with electron energy due to band nonparabolicity.
Average $g$ values are calculated summing over populated Landau and spin
levels properly accounting for the thermal distribution of electrons in
the band. It is shown that the spin splitting due to BIA in the presence
of magnetic field gives observable contributions to $g$ values. Our
calculations are in good
agreement with experiment in the temperature
range of 0 K to 300 K for CdTe and 0 K to 180 K for InP. The temperature
dependence of $g$ is stronger in CdTe than in InP due to different signs of
the band-edge $g$ values in the two materials. Good agreement between
the theory and experiment strongly indicates that the temperature dependence of spin $g$ factors is correctly explained.
In addition, we
discuss formulas for the energy dependence of spin $g$ factor due
to band nonparabolicity, which are liable to misinterpretation.

\end{abstract}

\maketitle

\section{\label{sec:level1}Introduction\protect\\ \lowercase{}}

Temperature dependence of the bulk electron spin $g$ factor in semiconductors is of interest both for scientific reasons as well as for possible spintronic applications. As far as the theory is concerned, a correct description of electron spin properties can test validity of the {\kp} theory for nonzero temperatures. As to the experiment, it is now possible to measure the electron spin $g$ value up to the room temperature using quantum beats in a time-resolved photoluminescence of spin states and related effects. The latter furnished a consistent experimental picture of  $g(T)$ for electrons in GaAs (see Ref. 1 and the references therein). It was shown that one can successfully describe the $g$ factor if, in the nonparabolic {\kp} theory, one takes the dilatation change of the energy gap, in agreement with previous theoretical predictions$^{2-4}$. On the other hand, one should account for the obvious fact that, as the temperature increases, more and more Landau and spin levels are populated by electrons. In nonparabolic conduction bands of III-V compounds the $g$ factors change with electron energy from their values at the band edge to the free electron value +2 at high energies$^{5, 6}$. Since $g$ depend on electron energy, one measures in reality their values averaged over all populated levels.

In our present paper we are concerned with temperature dependences of  the bulk electron $g$ factors in CdTe and InP. The temperature variations of $g$ in these materials have been measured but the theoretical descriptions are missing$^{7, 8}$. Ito et al$^8$ attributed almost all the temperature variation of $g$ in CdTe to the far-band contributions which was clearly not justified, as the authors themselves recognized. Thus, the subject can be regarded as controversial, as was previously the case for GaAs, see Refs. 7, 9, 10.

The band structure of CdTe and InP is similar to that of GaAs. In CdTe the $g$ value is negative at the band edge and it tends to +2 going through zero as the energy increases. In InP the $g$ value is positive at the band edge so that, tending to +2, its energy dependence is much slower. Thus we deal with two distinctly different cases. Another aspect of the present work is the spin splitting due to the bulk inversion asymmetry (BIA), called the Dresselhaus splitting$^{11}$. This splitting leads to a number of effects, also in the presence of an external magnetic field. In our description we account for the spin splitting due to BIA at finite magnetic fields and discuss this contribution. To our knowledge, this problem has not been considered before.

 Our work has three objectives. First, we describe the temperature dependence of the electron $g$ factors in CdTe and InP and compare the theory with existing experimental data. Second, we analyze the temperature dependences of $g$ values in the two cases and show that they result from the opposite signs of $g$ at the band edges of both materials. Third, we study the effect of Dresselhaus spin splitting at finite magnetic fields on the $g$ value. Finally, we discuss validity of a frequently employed  formula for the energy dependence of $g$ factor in III-V and II-VI compounds and indicate how it should be used.
Our paper is organized in the following way. In Section II we summerize the band structure calculations and indicate how the average $g$ values are computed. In Section III we describe our calculations for CdTe and compare them with experimental data, Section IV contains similar program for InP. In Section V we discuss our results and in Section VI we summerize them. In Appendix we consider the energy dependence of $g$ value, liable to misinterpretation.

\section{\label{sec:level1} THEORY\protect\\ \lowercase{}}

InP and CdTe are medium-gap semiconductors (MGS) and a three-level {\kp} description, successfully used for narrow gap
semiconductors$^{12, 13}$, is not adequate for describing their band structures. The reason is that in MGS the
fundamental gap $E_0$ between the $\Gamma^c_6$ and $\Gamma^v_8$ levels is not much
smaller than the gap $E_1$ between the $\Gamma^c_6$ level and the upper $\Gamma^c_7$ conduction level.
It has been demonstrated that an adequate way to treat the conduction band of MGS is
to use a five-level model (5LM), which is equivalent to 14 bands (including spin) in the {\kp} description (see
Refs. 14, 15 and the references therein). According to the
five-level model the spin $g$ value at the conduction band edge is [14],
$$
 g^*_0=2+\frac{2}{3}\left[E_{P_0}\left(\frac{1}{E_0}- \frac{1}{G_0}\right)+E_{P_1}\left(\frac{1}{G_1}-
 \frac{1}{E_1}\right)\right]+
$$
\begin{equation}
 -\frac{4{\overline{\Delta}}\sqrt{E_{P_0}E_{P_1}}}{9}\left(\frac{2}{{E_1}{G_0}}+
\frac{1}{{E_0}{G_1}}\right)+2C' \;\;,
\end{equation}
where $E_{P_0}$ = $2m_0P_0^2/\hbar^2$, $E_{P_1}$ = $2m_0P_1^2/\hbar^2$, $G_0$ = $E_0 + \Delta_0$ and $G_1 = E_1
+ \Delta_1$. The spin-orbit energies $\Delta_0$ and $\Delta_1$ relate to ($\Gamma^v_7$, $\Gamma^v_8$) and
($\Gamma^c_7$, $\Gamma^c_8$) levels, respectively, $\overline{\Delta}$ is the interband matrix element of the
spin orbit interaction between the ($\Gamma^v_7$, $\Gamma^v_8$) and ($\Gamma^c_7$, $\Gamma^c_8$) multiplets (see
[14, 16]), and $C'$ is due to far-band contributions. Figure 1 shows schematically the five-level {\kp} model
used in our calculations. For $\overline{\Delta}$ = 0 Eq. (1) reduces to the formula
given first by Hermann and Weisbuch$^{17}$. Calculating the electron energies away from the band edge one deals with
the effects of band's nonparabolicity and inversion asymmetry. In particular, an appearance of the matrix element $Q$
is due to the bulk inversion asymmetry. Since the 5LM for electrons in the presence of a
magnetic field and its use for magnetooptical properties of MGS was described in some details before$^{14, 15}$,
we only mention here the main elements of this approach. Thus the model includes exactly the $\Gamma^v_7$,
$\Gamma^v_8$, $\Gamma^c_6$, $\Gamma^c_7$ and $\Gamma^c_8$ levels at the center of the Brillouin zone and the
resulting {\kp} matrix has dimensions 14 $\times$ 14. There exist three nonvanishing interband matrix elements
of momentum: $P_0$, $P_1$ and $Q$.
If one takes Q = 0 and $k_z$ = 0 (where $\hbar k_z$ is the momentum along the magnetic field) the 14
$\times$ 14 initial matrix factorizes into two 7 $\times$ 7 matrices for the spin-up and spin-down states. These
matrices are soluble by envelope functions in the form of harmonic oscillator functions and the
eigenenergy problem for different Landau levels (LLs) $n$ reduces to diagonalization of 7 $\times$ 7
determinants.

Taking the asymmetric gauge for the vector potential \textbf{  \emph{A}} = [-$By$, 0, 0] one obtains the harmonic
oscillator functions in the form $exp(ik_x x +i k_z z) \Phi_n [(y-y_0)/L]$, where $y_0 = k_x L^2$. Here $L=(\hbar/eB)^{1/2}$
is the magnetic radius.
If the $Q$ element is included (it comes from an inversion asymmetry of MGS crystals)
the initial 14 $\times$ 14 matrix does not factorize and is not soluble in terms of a single column of
harmonic oscillator functions. Physically, this means that the resulting energy bands are not spherical. Since
the nonsphericity of the conduction bands in MGS is small, one can find the eigenenergies looking for the
envelope functions in terms of sums of harmonic oscillator functions (see Ref. 18). This leads to number determinants composed
of the fundamental 7 $\times$ 7 blocks on the diagonal coupled by nondiagonal parts involving Q and $k_z$ elements.
The eigenenergies are computed truncating the resulting big determinants. In our computations we used typically
112 $\times$ 112 determinants. All calculations were performed taking a magnetic field \textbf{ \emph{B}}
parallel to the [001] direction.

\begin{figure}
\includegraphics[scale=1.1,angle=0, bb = 50 25 182 300]{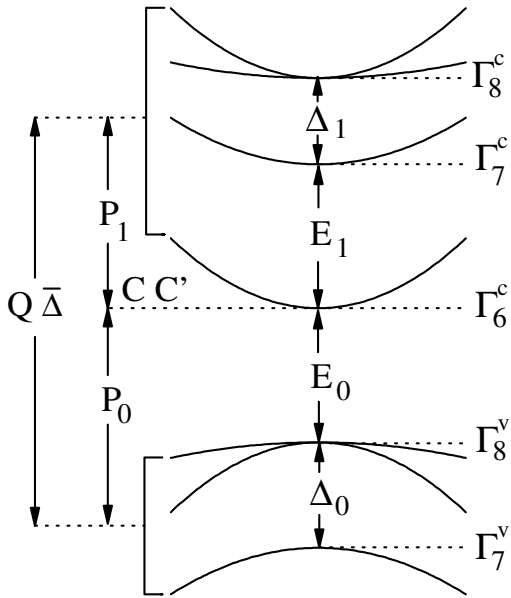}
\caption{\label{fig:epsart}{Five-level {\Pp} model of band structure in medium-gap semiconductors
CdTe and InP. The zero of energy is chosen at the $\Gamma^c_6$ edge. Interband
matrix elements of momentum $P_0$, $P_1$, $Q$ as well as interband matrix element of
spin-orbit interaction $\overline{\Delta}$ are indicated, $C$ and $C'$ symbolize
far-band contributions to the band-edge effective mass and spin $g$ factor,
respectively.}} \label{fig1th}
\end{figure}

Next, we consider average values of the spin $g$ factor measured as a function of temperature. The
measurements are usually done in relatively pure samples having low free electron densities. The electrons are
excited across the gap into the conduction band and into both spin states. The spin states are almost equally
populated and the circularly polarized light produces a well defined coherence between them. The excited
electrons quickly thermalize and are distributed among Landau levels (LLs) according to the lattice temperature without losing
their spin or phase. Then they interfere and quantum beats in the photoluminescence or other effects are
observed from many LLs. According to this picture the observed signal represents an average over the populated
levels. The electron thermal distribution over LLs determines their contribution to the average $g$ value.

We assume the $k_z$-dependence of electron energies in a simplified form
\begin{equation}
{\cal E}_{nk_z}^{\pm}={\cal E}_n^{\pm}+\frac{\hbar^2k^2_z}{2m^*_0} \;\;,
\end{equation}
where $n$ is the LL number, $\pm$ signs correspond to the two spin states, $k_z$ is the wavevector along the direction of
\textbf{\emph{B}}, and $m^*_0$ is the effective mass at the band edge. The description of energies ${\cal E}_n^{\pm}$ contains the intricacies of
the band structure mentioned above. The
spin $g$ value is defined as (in formulas we use $g^*$ symbol)
\begin{equation}
g^*=({\cal E}^+_{nk_z}-{\cal E}^-_{nk_z})/\mu_BB.
\end{equation}

An averaging procedure involves a summation over $n$ and integrations over $k_x$ and $k_z$. A simple calculation gives the
average value of $g^*$ in the form
\begin{figure}
\includegraphics[scale=0.65,angle=0, bb = 100 25 276 290]{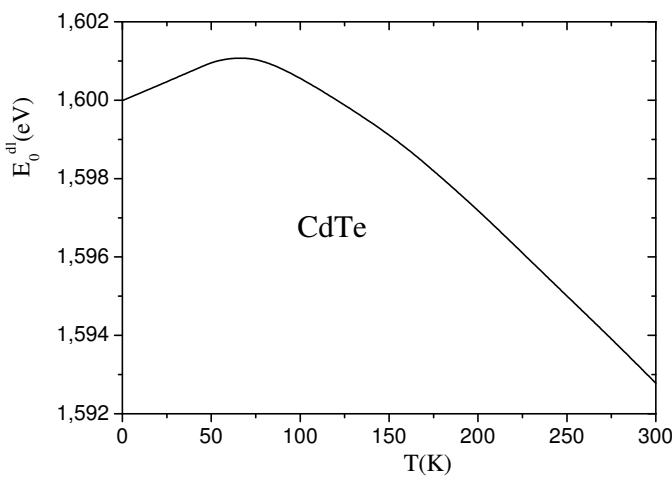}
\caption{\label{fig:epsart}{Dilatation gap $E_0 (T)$ in CdTe versus temperature calculated from
Eq.(7) with the use of experimental values of $D$, $\partial E_0/\partial P$ and $\alpha_{th}(T)$, see
Eq. (2).}} \label{fig2th}
\end{figure}

\begin{equation}
{\overline{g^*}(T)}=\frac{A}{C}\;\;,
\end{equation}
where
\begin{equation}
A=\sum^{\infty}_{n=0}\int^{\infty}_{{\cal E}^i_n} \frac{g^*_n({\cal E})f({\cal E,\zeta})}{({\cal E}-{\cal
E}^i_n)^{1/2}}d{\cal E}\;\;,
\end{equation}
and
\begin{equation}
C=\sum^{\infty}_{n=0}\int^{\infty}_{{\cal E}^i_n} \frac{f({\cal E,\zeta})}{({\cal E}-{\cal E}^i_n)^{1/2}}d{\cal
E} \;\;,
\end{equation}
in which the summation is over the LLs, $f({\cal E,\zeta})$ is the Fermi-Dirac distribution function, and the square roots come from
the integrations over $k_z$. The integrations begin from the lower of the two states ${\cal E}^i_n$ for each $n$, which can be either ${\cal
E}^+_n$ or ${\cal E}^-_n$ depending on the sign of the $g$ value.

The average $g$ value, as given by Eq. (4), is affected by the temperature in two opposite ways. As the temperature
$T$ increases and the absolute value of the fundamental gap $E_0$ decreases, the spin $g^*_0$ value at the band edge
decreases. On the other hand, with increasing T the
electrons populate higher LLs and band's nonparabolicity comes more and more into play. The latter is known to
make the $g$ value less negative (see Refs. 5, 6). Thus, as $T$ increases, the average $g^*$
decreases or increases depending on the relative strength of the two effects. We emphasize that we do not use in our
calculations Eq. (1), it is quoted only to make clear the dependence of $g^*_0$ on $E_0$ and other parameters.

\section{\label{sec:level1} C\lowercase{d}T\lowercase{e}\protect\\}

As mentioned above, it was demonstrated that the temperature change in the effective mass and the spin $g$ factor in a material
is governed by a \emph{dilatational} variation of energy gaps, of which the fundamental gap is of primary importance.
Thus one needs to determine the dilatational change in the fundamental gap due to temperature since the directly measured total
temperature change is due to both the dilatation of the crystal lattice and its vibrations (phonons).
We quote the determination of $E^{dl}_0(T)$ for CdTe since, to our knowledge, it has not been carried out before.
The dilatational change in the gap is given by
\begin{figure}
\includegraphics[scale=0.6,angle=0, bb = 140 25 282 320]{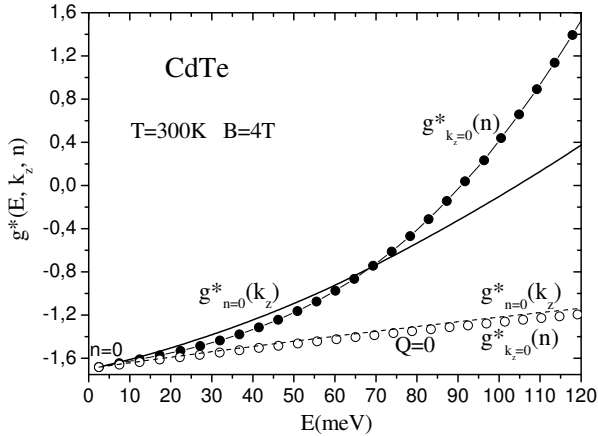}
\caption{\label{fig:epsart}{Calculated spin $g$ factor in bulk CdTe versus electron energy in the
conduction band. Full points are $g$ values for consecutive LLs and $k_z$ = 0,
solid line indicates $g$ value for $n = 0$ as a function of $k_z$, empty points and
dashed line: the same as above but neglecting bulk inversion asymmetry (
i.e. putting $Q$ = 0).}} \label{fig3th}
\end{figure}

\begin{equation}
\Delta E^{dl}_0(T)=-3D\left(\frac{\partial E_0}{\partial P}\right)_T \int^T_0 \alpha_{th}(T')dT' \;\;,
\end{equation}
where $D$ is the bulk modulus, $\partial E_0/\partial P$ is the pressure-induced gap shift, and $\alpha_{th}(T)$ is the linear
thermal expansion coefficient (see also Ref. 19). The quantities $D$ and $\partial E_0/\partial P$ are directly
measurable, for CdTe there is $\partial E_0/\partial P$ = 0.08 eV/GPa$^{20, 21}$ (see also Refs. 22-26) and $D$ = 42 GPa$^{26}$.
The linear thermal expansion coefficient $\alpha_{th}(T)$ was calculated$^{28}$ and measured$^{29}$.
Using these results we performed the integration indicated in Eq. (7) and obtained the dilatational variation of the
fundamental gap in CdTe shown in Fig. 2. It is seen that E$^{dl}_0(T)$ goes through a maximum at $T\approx$ 65
K but, all in all, the temperature variation between 0 K and 300 K is rather small.

Knowing E$^{dl}_0(T)$ and assuming that other gaps, momentum matrix elements and far-bands contributions do not depend on the
temperature, one can perform the band structure calculations outlined above if the band parameters are known. For CdTe we take
the following parameter values at $T$ = 0: $E_{P_0}$ = 21.07 eV, $E_{P_1}$ = 5.1 eV, $E_Q$ = 13.29 eV (see Ref. 30), $E_0$ = -1.6 eV$^{31}$, $\Delta_0$ = -0.95 eV$^{32, 33}$, $E_1$ = 3.76 eV$^{34, 35}$, $\Delta_1$ = 0.27 eV$^{30}$, $\overline{\Delta}$ = -0.19 eV$^{30}$, $C$ = -0.5355 and $C'$=-0.0129. The far-band contributions are taken to obtain at $T$ = 0 the band edge values $m^*_0$ = 0.093 $m_0$$^{36}$ and $g^*_0$ = -1.66$^{9}$. The zero of energy is chosen at the $\Gamma^c_6$ edge, see Fig. 1, so the energies above are positive while the energies below are negative.

Figure 3 shows the result of our intermediate calculations for CdTe, given as example. The calculations are performed for fixed
values of $T$ and $B$. In order to investigate the effect of BIA on
the $g$ value at finite $B$ we carried out the computations in two versions: 1) using the full 5LM, i. e. including
the matrix element $Q$ which, as mentioned above, incorporates BIA (full points and solid line); 2) putting $Q$ = 0,
i. e. neglecting the effect of BIA (empty points and dashed line). The full points in Fig. 3 indicate calculated spin $g$ values for consecutive LLs (at $k_z$ = 0) beginning with LL $n$ = 0. It is seen that the $g$ value increases with the LL number $n$ (or, equivalently, the energy) due to band's nonparabolicity. The solid line shows the calculated $g$ for $n$ = 0 as a function of $k_z$ (or, equivalently, the energy). It can be seen that, for lower LL numbers $n$, the $g$ value behaves very similarly for $n$-dependence and $k_z$-dependence.
As argued by the present authors$^{37}$, the above dependences should be identical within the description by the three-level model, see also Appendix. The empty points and dashed line show the corresponding quantities calculated with $Q$ = 0. Here both $g^*(n, k_z=0)$ and $g^*(n=0, k_z)$ are practically the same.
The differences between the full points (and solid line) and the empty points (and dashed line) are directly due to
the effect of BIA.

It is well known that, in the absence of magnetic field, the spin splitting due to BIA is $\Delta {\cal E} = 2\gamma[k^2(k_x^2 k_y^2 + k_x^2 k_z^2 +k_y^2 k_z^2)-9k_x^2 k_y^2 k_z^2]^{1/2}$.
For the presence of $B$ we can not give an analytical expression for the spin splitting due to BIA. However, one can qualitatively say that, at $B \ne 0$, $k_x$ and $k_y$ components are replaced by $(nB)^{1/2}$ or $[(n+1)B]^{1/2}$ terms, while the $k_z$ component along the magnetic field remains the same. For $n = 0$ the "transverse" components are small (or zero) and for ${\cal E} \approx 0$ the longitudinal component $k_z \approx 0$, so that BIA gives almost no contribution, which agrees with the results shown in Fig. 3. The increasing energy ${\cal E}$ corresponds to the increase of $n$ or $k_z$ (or both), so the contribution of BIA grows. This is reflected by an increasing difference of the results for $Q \ne 0$ and $Q$ = 0 shown in Fig. 3.

We follow the similarity of $g^*$($n$ =const, $k_z$) and $g^*$($n$, $k_z$ = const) dependences shown in Fig. 3\emph{ assuming} $g^*$($n$ =const, $k_z$) to be equal to to $g^*$($n$, $k_z$ = const) in the summation of LLs and integration over $k_z$, see Eqs. (5) and (6). This approximation is in fact quite good since the region of high energies: ${\cal E} \ge$ 90 meV for CdTe (see Fig. 3), where this approximation begins to break down, is weakly populated by electrons and it gives only small contribution to the average $g$ value.

\begin{figure}
\includegraphics[scale=0.6,angle=0, bb = 140 25 282 320]{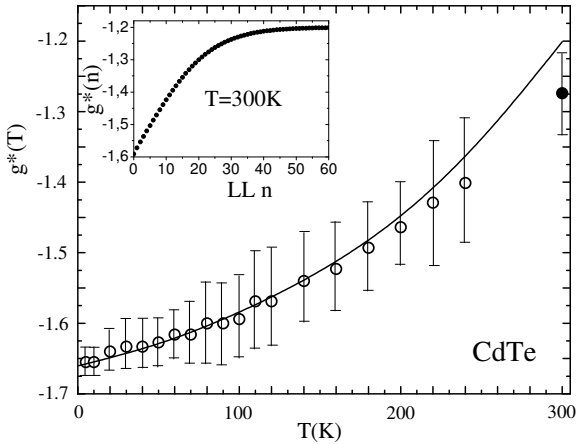}
\caption{\label{fig:epsart}{Spin $g$ factor in bulk CdTe versus temperature. Empty points:
experimental data of Ref. 7, full point: data of Ref. 8. Solid line:
theoretical average $g$ factor calculated according to Eq. (4). Inset
shows how consecutive LLs contribute to the average $g$ value at $T = 300$ K.}} \label{fig4th}
\end{figure}

Using the above assumption we performed the summation and integration indicated in Eqs. (5) and (6) and determined the average value of $g$ according to Eq. (4).
Our final results for CdTe are shown in Fig. 4. It can be seen that the experimental values of Oestreich et al$^{7}$ (lower temperatures) and that of Ito et al$^{8}$ (room temperature) are described very well. For temperatures above 200 K the theory is slightly higher then experiment. We conclude that the main reason behind the observed increase of the spin $g$ value with temperature is that, as the temperature increases, higher Landau and spin levels contribute to the everage $g$ value. Since the band structure predicts an increase of $g$ with growing energy (see also the discussion in Appendix) the resulting average $g$ becomes higher (less negative). This mechanism is well illustrated in the inset of Fig. 4.

\section{\label{sec:level1} I\lowercase{n}P\protect\\}

InP is a medium-gap semiconductor similar in many respects to GaAs. However, the electron spin $g$ factor at the band edge of InP is positive, in contrast to InSb, InAs, GaSb and GaAs$^{17}$. Since, as we mentioned above, at high electron energies $g$ tends to +2 (if one neglects the effects of BIA) there remains not much room for the energy variation of $g$ value.

The dilatational change in the fundamental energy gap of InP was estimated by Hazama et al$^{4}$. However, we revise this  estimation because a part
of the procedure adopted in Ref. 4 was based on calculations, while in our approach we use exclusively experimental information.
  To determine the dilatation gap of InP we use again formula (7). We take
$D$ = 71 GPa$^{38, 39}$, and $dE_0/dP$ =0.084 eV/GPa$^{40}$.
As to the function $\alpha_{th}(T)$, it was measured by various authors, see Refs. 41-43. We follow the measurements of Haruna et al$^{41}$, which basically agree with those of other authors but are more complete. In the data of Ref. 41 we correct the point at T = 8 K (since the given value $\alpha_{th}(8 K)$ has the wrong sign) by interpolating between the measured values at 0 K and 15 K. The complete function $\alpha_{th}(T)$ is then
used for the numerical integration indicated in Eq. (7). Our final results for $E^{dl}_0(T)$  are shown in Fig. 5; the obtained variation in the dilatation gap
is noticeably smaller than that given in Ref. 4. We use the results indicated in Fig. 5 in our further procedure.

To perform the band structure calculations we take the following band parameters$^{44}$: $E_{P_0}$ = 20.93 eV, $E_{P_1}$ = 0.165 eV, $E_Q$ = 15.56 eV, $E_0$ = -1.423 eV, $\Delta_0$ = -0.108 eV, $E_1$ = 3.297 eV, $\Delta_1$ = 0.201 eV, $\overline{\Delta}$ = 0.08733 eV$^{45}$, $C$ = -2.467 and $C'$ = -0.08045. The far-band contributions are taken to obtain at $T$ = 0 the band edge values $m^*_0$ = 0.07927$m_0$$^{36}$ and $g^*_0$ = +1.204. Our band structure calculations are performed similarly to those described above for CdTe.

\begin{figure}
\includegraphics[scale=0.6,angle=0, bb = 140 25 282 320]{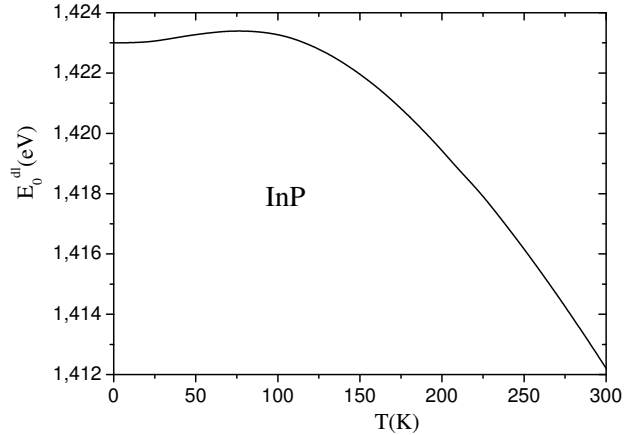}
\caption{\label{fig:epsart}{The same as in Fig. 2 but for InP.}} \label{fig5th}
\end{figure}

In Fig. 6 we show intermediate calculated results for $g^* (n, k_z=0)$ and $g^*(n=0, k_z)$ according to the 5LM in two versions: 1) including the matrix element $Q$, i.e. including BIA (full points and solid line); 2) putting $Q$ = 0, i.e. neglecting BIA (empty points and dashed line). Again, as discussed above, in the region of energies ${\cal E} \le$ 190 meV the two dependences are very similar (for $Q$ = 0 they practically coincide for all energies) so that, when performing the integration over $k_z$ in Eq. (5) we \emph{assume} that the $k_z$-dependence of $g$ value for each LL is the same as that given by the black points. As far as the contribution of BIA to the spin splitting is concerned, the picture is similar to that for CdTe: at low energies the contribution of BIA vanishes but it grows with the energy. Clearly, in our final calculations we do take into account the effect of BIA.

Figure 7 shows our final results for the temperature dependence of the average $g$ value in InP, computed according to Eqs.(4)-(6). It is seen that the available experimental data of Oestreich et al [7] are described very well. Similarly to CdTe, the main reason for the increase of $g$ value with the temperature is that, as $T$ grows, more Landau and spin levels in the nonparabolic conduction band of InP are populated with electrons. On the other hand, the temperature variation of average $g$ in InP is considerably weaker than that in CdTe for reasons given above.

\begin{figure}
\includegraphics[scale=0.6,angle=0, bb = 140 25 282 320]{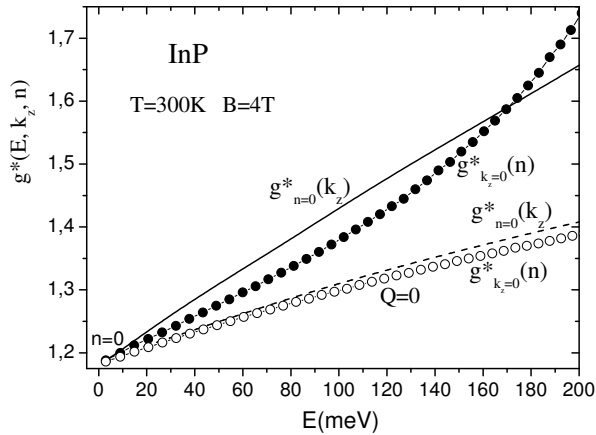}
\caption{\label{fig:epsart}{The same as in Fig. 3 but for InP.}} \label{fig6th}
\end{figure}

\section{\label{sec:level1} DISCUSSION\protect\\}

One should bear in mind that the $g$ values measured by various authors do not agree too well with each other. For CdTe, Meyer et al$^{46}$ used the spin resonance to measure $g$ values in the temperature range of 4.2 K $\le T \le$ 66 K and obtained the data which can be described by the linear dependence $g^*(T)$ = -1.682 + 2.97$\times$10$^{-4}$ T. All these values are somewhat lower than those shown in Fig. 4. Sprinzl et al$^{47}$  used the spin quantum beats to measure the $g$ values that agree with those shown in Fig. 4 at low temperatures but are considerably lower near room temperature.
We believe that the data shown in Fig. 4 are reliable since the lower temperature results of Oestreich et al$^{7}$ are consistent with the result of Ito et al$^8$ for $T$ = 300 K.

\begin{figure}
\includegraphics[scale=0.6,angle=0, bb = 140 25 282 320]{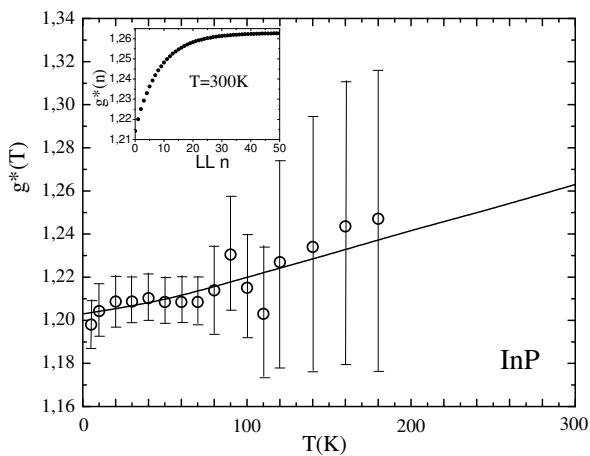}
\caption{\label{fig:epsart}{Spin $g$ factor in bulk InP versus temperature. Empty points:
experimental data of Ref. 7. Solid line:
theoretical average $g$ factor calculated according to Eq. (4). Inset
shows how consecutive LLs contribute to the average $g$ value at $T = 300$ K.
}} \label{fig7th}
\end{figure}

As for InP, we note that the low-temperature result of Weisbuch and Hermann$^{48}$ $g^*_0$ = +1.26, obtained with the use of spin resonance, is distinctly higher than the values shown in Fig. 7. However, such differences between measured $g$ values are not unusual, see Fig. 4 of Ref. 1 for GaAs. We were unable to find experimental measurements of the $g$ factor in InP for temperatures above 200 K.

As follows from Figs. 3 and 6, the $g$ values calculated taking $Q \ne$ 0, i.e. using the complete 5LM description, are for higher electron energies considerably higher than those obtained taking $Q$ = 0. This means that the agreements between experiment and theory shown in Figs. 4 and 7 give clear indications that the influence of bulk inversion asymmetry on the spin splitting at $B \ne$ 0 is not negligible in CdTe and InP. This evidence is still somewhat indirect since the average values are given by sums over many LLs and  integrations over $k_z$. However, as each term of the sum over LLs and the integration over $k_z$ is affected by BIA, the evidence for its non-negligible effect seems convincing. One could verify more directly the effect of BIA in the presence of a magnetic field by investigating at low temperatures an anisotropy of the spin splitting with respect to the orientation of a magnetic field. It should be mentioned that the effect of BIA on the $g$ values of heavy holes in GaAs quantum wells was recently investigated by Kubisa et al$^{49}$.

Finally, we believe that the distinctly different temperature dependences of spin $g$ values in CdTe and InP, both experimental and theoretical, are significant. The increase of $g$ in CdTe between $T \approx$ 0 and $T$ = 180 K is $\Delta g^* \approx$ 0.16, while in InP the increase in the same temperature range is $\Delta g^* \approx$ 0.049. As we remarked above, this difference is due to different energy dependences of $g$ factors in the two materials because at high energies the $g$ factor should reach the free-electron value of +2 (if one neglects BIA), while the initial band-edge value in CdTe is $g^*_0 \approx$ -1.66 and in InP it is $g^*_0 \approx$ +1.20. In consequence, $g^*(\cal{E})$ dependence in InP is distinctly weaker than that in CdTe. The above reasoning is confirmed experimentally which strongly indicates that our interpretation, as described above, is correct. We emphasize this conclusion since the reason for the temperature dependence of the spin $g$ factor remains to be a matter of controversy, cf. Ref. 10.

\section{\label{sec:level1}SUMMARY\protect\\ \lowercase{}}

Dilatational changes in the fundamental energy gaps of CdTe and InP are
determined from available experimental data in order to use them in
calculations of band structures at nonzero temperatures. The five-level {\Pp}
model is employed to compute the band structures of these medium-gap
semiconductors in the presence of a magnetic field. In particular, energy
dependences of the electronic spin $g$ factors due to band nonparabolicities
are obtained. Next, average $g$ values are calculated for different
temperatures summing over populated Landau  and spin levels and
integrating over longitudinal momentum $\hbar k_z$. The increase of $g$'s with
growing temperature are almost exclusively due to the population of
higher Landau and spin levels in nonparabolic conduction bands. The
calculated $g^*(T)$ dependences are compared with available experimental
values and good agreement between experiment and theory is obtained for
both materials. The temperature increase of spin $g$ factor is stronger
in CdTe than in InP, which is related to the negative band-edge value of $g$
in CdTe and the positive one in InP. It is shown that the bulk inversion
asymmetry gives observable contributions to the spin splittings in
the presence of a magnetic field. Frequently used formulas for the
energy-dependent spin $g$ factors in nonparabolic conduction bands of III-V
compounds are discussed.

\appendix*
\section{}

Since our work is closely related to dependence of the electron spin $g$ factor on the electron energy ${\cal E}$ in III-V and some II-VI semiconductor compounds, we discuss here the use of an often employed formula for $g^*({\cal E})$. To the best of our knowledge, the first description of $g^*({\cal E})$ in narrow-gap III-V compounds was given in 1961 by Lax, Mavroides, Zeiger, and Keyes$^{5}$ (LMZK) and a different but equivalent formula was derived in 1963 by Zawadzki$^{6}$. The formula of LMZK, although old, is still used in the literature (see for example Refs. 8, 50-52) so it merits a discussion.

It is important to see how the formula is derived. The underlying formulation is due to Bowers and Yafet$^{12}$, see also Ref. 6. The three-level {\Pp} description (see Refs. 12, 53) takes into account the $\Gamma^c_6$ conduction level and $\Gamma^v_8$, $\Gamma^v_7$ valence
levels. The resulting {\Pp} 8$\times$8 Hamiltonian is solved in terms of harmonic oscillator
functions neglecting small free-electron terms. Final equation for the energies
is obtained in the form
$$
{\cal E}({\cal E}-E_0)({\cal E}-E_0-\Delta_0)+
$$
\begin{equation}
-P^2_0[s(2n+1)+k^2_z]({\cal E}-E_0-\frac{2}{3}\Delta_0) \mp\frac{1}{3}P^2_0\Delta_0s = 0\;\;,
\end{equation}
where $s = eB/\hbar$ and other symbols have been defined above. For specified $B, n, k_z$ and the spin sign $\pm$, Eq. (A.1) represents a cubic equation for the energy ${\cal E}(n, k_z, \pm)$. In order to proceed further, we divide Eq. (A.1) by $({\cal E}-E_0)({\cal E}-E_0-\Delta_0)$ and after a simple algebraic manipulation obtain for the conduction band
\begin{equation}
{\cal E}^{\pm}_{nk_z}=
\frac{\hbar e B}{m^*({\cal E}^{\pm}_{nk_z})}(n+\frac{1}{2})+\frac{\hbar^2 k_z^2}{2m^*({\cal E}^{\pm}_{nk_z})} \pm\frac{\mu_BB}{2}g^*( {\cal E}^{\pm}_{nk_z})
\end{equation}
where
\begin{equation}
\frac{m_0}{m^*({\cal E}^{\pm}_{nk_z})}=1-\frac{E_{P_0}}{3}
\left(\frac{2}{{\tilde E_0}^{\pm}}+ \frac{1}{{\tilde G_0}^{\pm}}\right);\;\;\,
\end{equation}

\begin{equation}
 g^*({\cal E}^{\pm}_{nk_z})=2+\frac{2E_{P_0}}{3}\left(\frac{1}{{\tilde E_0}^{\pm}}- \frac{1}{{\tilde G_0}^{\pm}}\right)
\end{equation}
in which
\begin{equation}
{\tilde E_0}^{\pm} = E_0 - {\cal{E}}^{\pm}_{nk_z}\;\;,
\end{equation}
\begin{equation}
{\tilde G_0}^{\pm} = E_0 + \Delta_0 - {\cal{E}}^{\pm}_{nk_z}\;\;.
\end{equation}
The additive terms +1 in Eq. (A.3) and +2 in Eq. (A.4) result from the free-electron contributions in the {\Pp} theory, which were neglected in the original Bowers and Yafet treatment$^{12}$, cf. Ref. 14. Equations (A.2), (A.3), (A.4) amount to the LMZK formulas in which band's nonparabolicity enters via the energy dependence of $m^*$ and $g^*$ (LMZK formulas do not contain the free electron terms).

The problem with the above formulation is that the energy  ${\cal E}^{\pm}_{n k_z}$ on the LHS of Eq. (A.2) is the same as that entering $m^*({\cal E}^{\pm}_{n k_z})$ and $g^*({\cal E}^{\pm}_{n k_z})$ on the RHS. Suppose one fixes $B, n$ and $k_z$ and tries to calculate the spin splitting. It is clear that it is not enough to take the energy difference for the plus and minus signs in Eq. (A.2) and calculate the energy on LHS because both the mass $m^*({\cal E}^{\pm}_{n k_z})$ and the spin factor $g^*({\cal E}^{\pm}_{n k_z})$ have different values for different signs of the spin. In particular, if $n$ or $k_z$ are large, the orbital term in Eq. (A.2) is considerably different for the opposite spins which should be necessarily taken into account. In fact, the form given by LMZK and used in the literature is inconvenient since the energy ${\cal E}^{\pm}_{n k_z}$ should be determined self-consistently. It is considerably simpler to use directly Eq. (A.1) and solve for the energies. However, in most applications one simply employs formula (A.4) for the energy dependence of $g^*({\cal E})$ forgetting the orbital term and the fact that in a nonparabolic band $g^*({\cal E}^{\pm}_{n k_z})$ is different for each spin direction.

Finally, we note that a formulation analogous to that given by LMZK for 3LM has been derived by the present authors for GaAs-type medium gap semiconductors with the use of 5LM, see Refs. 37, 54. The result is that Eq. (A.2) is still valid but the formulas for $m^*({\cal E}^{\pm}_{nk_z})$ and $g^*( {\cal E}^{\pm}_{nk_z})$ become

$$
\frac{m_0}{m^*({\cal E}^{\pm}_{nk_z})}=1+C-\frac{1}{3}\left[E_{P_0}\left(\frac{2}{{\tilde E_0}^{\pm}}+ \frac{1}{{\tilde G_0}^{\pm}}\right)+E_{P_1}\left(\frac{2}{{\tilde G_1}^{\pm}}+ \frac{1}{{\tilde E_1}^{\pm}}\right)\right]
$$
\begin{equation}
+\frac{4{\overline{\Delta}}\sqrt{E_{P_0}E_{P_1}}}{3}\left(\frac{1}{{{\tilde E_1}^{\pm}}{{\tilde G_0}^{\pm}}}- \frac{1}{{{\tilde E_0}^{\pm}}{{\tilde G_1}^{\pm}}}\right) \;\;,
\end{equation}
and
$$
 g^*({\cal E}^{\pm}_{nk_z})=2+2C'+\frac{2}{3}\left[E_{P_0}\left(\frac{1}{\tilde E_0}- \frac{1}{{\tilde G_0}^{\pm}}\right)+E_{P_1}\left(\frac{1}{{\tilde G_1}^{\pm}}- \frac{1}{{\tilde E_1}^{\pm}}\right)\right]
$$
\begin{equation}
 -\frac{4{\overline{\Delta}}\sqrt{E_{P_0}E_{P_1}}}{9}\left(\frac{2}{{\tilde E_1}^{\pm}{\tilde G_0}^{\pm}}+
\frac{1}{{{\tilde E_0}^{\pm}}{{\tilde G_1}^{\pm}}}\right) \;\;,
\end{equation}
where ${\tilde E_0}^{\pm}$ and ${\tilde G_0}^{\pm}$ are still given by Eqs. (A.5) and (A.6), respectively, and
\begin{equation}
{\tilde E_1}^{\pm} = E_1 - {\cal E}^{\pm}_{nk_z}\;\;,
\end{equation}

\begin{equation}
{\tilde G_1}^{\pm} = E_1 + \Delta_1 - {\cal E}^{\pm}_{nk_z}\;\;.
\end{equation}
The terms proportional to $E_{P_1}$ are related to the interband $\Gamma^c_6$-$\Gamma^c_8$ and $\Gamma^c_6$-$\Gamma^c_7$ interactions, see Fig. 1. On both sides of Eq. (A.2) the energy is ${\cal E}^{\pm}_{n k_z}$ and, for given $B$, $n$, $k_z$ and $\pm$, this energy should be determined in a self-consistent way. The above formulas do not contain the element $Q$, i.e. they neglect the effects of bulk inversion asymmetry.


\begin{thebibliography}
{99}\label{sec:TeXbooks}
\bibitem{pp1}  W. Zawadzki and P. Pfeffer, R. Bratschitsch, Z. Chen, and S. T. Cundiff, B. N. Murdin, C. R. Pidgeon, Phys. Rev. B \textbf{78}, 245203 (2008).

\bibitem{pp2} H. Ehrenreich, J. Phys. Chem. Solids \textbf{2}, 131 (1959).

\bibitem{pp3} R. A. Stradling and R. A. Wood, J. Phys. C \textbf{3}, L94 (1970).

\bibitem{pp4} H. Hazama, T. Sugimasa, T. Imachi, and C. Hamaguchi, J. Phys. Soc. Jpn. \textbf{55}, 1282 (1986).

\bibitem{pp5} B. Lax, J. G. Mavroides, H. J. Zeiger, and R. J. Keyes, Phys. Rev. \textbf{122}, 31 (1961).

\bibitem{pp6} W. Zawadzki, Physics Lett. {\bf 4}, 190 (1963).

\bibitem{pp7} M. Oestreich, S. Hallstein, A. P. Heberle, K. Eberl, E. Bauser, and W. W. Ruhle, Phys. Rev.
B \textbf{53}, 7911 (1996).

\bibitem{pp8} T. Ito, W. Shichi, Y. Okami, M. Ichida, H. Gotoh, H. Kamada, and H. Ando, phys. stat. sol. (c) \textbf{6}, 319 (2009).

\bibitem{pp9} M. Oestreich and W. W. Ruhle, Phys. Rev. Lett. \textbf{74}, 2315
(1995).

\bibitem{pp10} J. Huebner, S. Dohrmann, D. Hagele, and M. Oestreich, Phys. Rev. B \textbf{79}, 193307 (2009).

\bibitem{pp11} G. Dresselhaus, Phys. Rev. \textbf{100}, 580 (1955).

\bibitem{pp12} R. Bowers and Y. Yafet, Phys. Rev. {\bf 115}, 1165 (1959).

\bibitem{pp13} C. R. Pidgeon and R. N. Brown, Phys. Rev. \textbf{146}, 575 (1966).

\bibitem{pp14} P. Pfeffer and W. Zawadzki, Phys. Rev. B \textbf{41}, 1561
(1990).

\bibitem{pp15} P. Pfeffer and W. Zawadzki, Phys. Rev. B \textbf{53}, 12813 (1996).

\bibitem{pp16} F.H.Pollak, C.W.Higginbotham, and M.Cardona,
J. Phys. Soc. Jpn. Suppl. {\bf 21}, 20 (1966).

\bibitem{pp17} C. Hermann and C. Weisbuch, Phys. Rev. B \textbf{15}, 823 (1977).

\bibitem{pp18} V. Evtuhov, Phys. Rev. {\bf 125}, 1869 (1962).

\bibitem{pp19} P. Lautenschlager, M. Garriga, S. Logothetidis, and M. Cardona, Phys. Rev. B \textbf{35}, 9174
(1987).

\bibitem{pp20} J. R. Mei and V. Lemos, Solid State Commun. \textbf{52}, 785 (1984).

\bibitem{pp21} D. L. Camphausen, G. A. Nevill Connell and W. Paul, Phys. Rev. Lett. {\bf 26}, 184 (1971).

\bibitem{pp22} G. A. Babonas, R. A. Bendoryus, and A. Yu. Shileika, Soviet Physics Semiconductors \textbf{5}, 392 (1971).

\bibitem{pp23} W. Shan, S. C. Shen and H. R. Zhu, Solid State Commun. \textbf{55}, 475 (1985).

\bibitem{pp24} M. Prakash, M. Chandrasekhar, and H. R. Chandrasekhar, Phys. Rev. B \textbf{42}, 3586 (1990).

\bibitem{pp25} H. M. Cheong, J. H. Burnett and W. Paul, Solid State Commun. \textbf{77}, 565 (1991).

\bibitem{pp26} J. Gonzalez, F. V. Perez, E. Moya and J. C. Chervin, J. Phys. Chem. Solids \textbf{56}, 335
(1995).

\bibitem{pp27} K. Strossner, S. Ves, W. Dieterich, W. Gebhardt and M. Cardona, Solid State Commun. \textbf{56}, 563 (1985).

\bibitem{pp28} D. Bagot, R. Granger, and S. Roland, phys. stat. sol. (b) \textbf{177}, 295 (1993).

\bibitem{pp29} G. K. White, J. G. Collins, J. A. Birch, and T. F. Smith, J. Phys. C \textbf{13}, 1649
(1980).

\bibitem{pp30} W. Willatzen, M. Cardona, N. E. Christensen, Phys. Rev. B \textbf{51}, 17992 (1995).

\bibitem{pp31} M. Nawrocki, A. Twardowski, phys. stat. sol. (b) \textbf{97}, K61 (1997).

\bibitem{pp32} A. Twardowski, E. Rokita, and J. A. Gaj, Sol. State. Communn. \textbf{36}, 927 (1980).

\bibitem{pp33} D. Niles, H. Hochst, Phys. Rev. B \textbf{43}, 1492 (1991).

\bibitem{pp34} V. V. Sobolev, O. G. Maksimova, S. G. Kroitoru, phys. stat. sol. (b) \textbf{103}, 499 (1981).

\bibitem{pp35} K. Boujdaria and O. Zitouni, Solid State Commun. \textbf{129}, 205 (2004).

\bibitem{pp36} E. Molva, Le Si Dang, Phys. Rev. B \textbf{27}, 6222 (1983).

\bibitem{pp37} P. Pfeffer and W. Zawadzki, Phys. Rev. B \textbf{74}, 115309 (2006).

\bibitem{pp38} V. M. Glazov, K. Davletov, A. Ya. Nashelskii and M. M. Mamedov, Zh. Fiz. Khim. \textbf{51}, 10, 2558 (1977). In Russian.

\bibitem{pp39} D. N. Nichols, D. S. Rimai, and J. Sladek, Solid St. Commun. \textbf{36}, 667 (1980).

\bibitem{pp40} H. Muller, G. Trommer, M. Cardona and P. Vogel, Phys. Rev. B \textbf{21}, 4879 (1980).

\bibitem{pp41} K. Haruna, H. Maeta, K. Ohashi and T. Koike, J. Phys. C \textbf{20}, 5275 (1987).

\bibitem{pp42} P. Deus, H. A. Schneider, U. Voland, and K. Stiehler, phys. stat. sol. (a) \textbf{103}, 443 (1987).

\bibitem{pp43} N. N. Sirota and A. A. Sidorov, Dokl. Akad. Nauk SSSR \textbf{284}, 1111 (1985). In Russian.

\bibitem{pp44} M. A. Hopkins, R. J. Nicholas, P. Pfeffer, W. Zawadzki,
D. Gauthier, J. C. Portal, and M. A. DiForte-Poisson, Semicond. Sci. Technol.
 {\bf 2}, 568 (1987).

\bibitem{pp45} I. Gorczyca, P. Pfeffer, and W. Zawadzki, Semicond. Sci. Technol.
 {\bf 6}, 963 (1991).

\bibitem{pp46} B. K. Meyer, A. Hofstaetter, U. Leib, D.M. Hofmann J. Crys. Growth \textbf{184-185}, 1118 (1998).

\bibitem{pp47} D. Sprinzl, P. Horodyska, N. Tesarova, E. Rozkotova, E. Belas, R. Grill, P. Maly, and P. Nemec, ArXiv 1001.0869.

\bibitem{pp48} C. Weisbuch and C. Hermann, Solid State Commun. \textbf{16}, 659 (1975).

\bibitem{pp49} M. Kubisa, K. Ryczko, and J. Misiewicz, Phys. Rev. B \textbf{83}, 195324 (2011).

\bibitem{pp50} A. A. Kiselev, E. L. Ivchenko, and U. Roessler, Phys. Rev. B \textbf{58}, 16353 (1998).

\bibitem{pp51} I. A. Yugova, A. Greilich, D. R. Yakovlev, A. A. Kiselev, M. Bayer, V. V. Petrov, Yu. K. Dolgikh, D. Reuter, and A. D. Wieck,
Phys. Rev. B \textbf{75}, 245302 (2007)

\bibitem{pp52} K. L. Litvinenko, L. Nikzad, C. R. Pidgeon, J. Allam, L. F. Cohen, T. Ashley, M. Emeny, W.
Zawadzki, and B. N. Murdin, Phys. Rev. B \textbf{77}, 033204 (2008).

\bibitem{pp53} W. Zawadzki, in \emph{Narrow Gap Semiconductors, Physics and Applications}, edited by W. Zawadzki,
(Springer, Berlin, 1980), p. 85.

\bibitem{pp54} P. Pfeffer and W. Zawadzki, Phys. Rev. B \textbf{74}, 233303 (2006).

\end{thebibliography}
\end{document}